\documentclass[aps,prl,onecolumn,superscriptaddress,preprintnumbers,showpacs,floatfix]{revtex4}
\usepackage{graphics,colordvi}
\def\eq#1\en{\begin{equation} #1 \end{equation}}

\def\eqa#1\ena{\begin{eqnarray} #1 \end{eqnarray}}

\usepackage{graphicx,color}

\usepackage{amsmath,amssymb}

\makeatletter
\newcommand{\fmslash}[2][0mu]{%
  \mathchoice
    {\fmsl@sh\displaystyle{#1}{#2}}%
    {\fmsl@sh\textstyle{#1}{#2}}%
    {\fmsl@sh\scriptstyle{#1}{#2}}%
    {\fmsl@sh\scriptscriptstyle{#1}{#2}}}
\newcommand{\fmsl@sh}[3]{%
  \m@th\ooalign{$\hfil#1\mkern#2/\hfil$\crcr$#1#3$}}

\newcommand{\tr}{\hbox{tr}}

\begin{document}

\title{Noncommutative geometry in quantum field theory and the 
cosmogenic neutrino physics at the extreme energies}


\author{Josip Trampeti\'{c}}
\affiliation{Physics Division, Rudjer Bo\v skovi\' c Institute, Zagreb, Croatia}



\begin{abstract}
Analysis of the covariant $\theta$-exact noncommutative (NC)
gauge field theory (GFT), inspired by high energy cosmic rays experiments, is performed in the framework of the inelastic neutrino-nucleon scatterings. Next we have have found neutrino two-point function and shows a closed form decoupled from the hard ultraviolet (UV) divergent term, from softened ultraviolet/infrared (UV/IR) mixing term, and from the finite terms as well. For a certain choice of the noncommutative parameter $\theta$ which preserves unitarity, problematic UV divergent and UV/IR mixing terms vanish. Non-perturbative modifications of the neutrino dispersion relations are assymptotically independent of the scale of noncommutativity in both, the low and high energy  limits and may allow superluminal propagation.
\end{abstract}

\pacs{11.10.Nx; 13.15.+g; 13.60.Hb; 98.70.Sa}
\maketitle


\section{Introduction}

The idea of noncommutative coordinates is not new and it was for the first time proposed by Heisenberg in 1930 in his letters to Ehrenfest \cite{Ehrenfest} and Peierls \cite{Peierls}. At that time Heisenberg could not formulate this idea mathematically.
Still, there was a hope that uncertainty relations for coordinates might provide a
natural cut-off for divergent integrals of Quantum Field Theory (QFT). The idea was propagated and in 1943
Snyder published a paper on "Quantized Space Time" \cite{Snyder:1946qz}. Pauli in a letter to Bohr mentioned this work to be mathematically ingenious but he rejected it for physical reasons. Later, due to the extreme success of renormalization of Quantum Electrodynamics (QED), Quantum Chromodynamics (QCD) and/or Standard Model (SM), the idea of possible Noncommutaive Field Theory (NCFT) was ignored for a long time. However at the beginning of 90's Noncommutative Geometry was developed, see Connes \cite{Connes} and Madore \cite{Madore} and references there. Research on divergences in a field theory on quantum spaces started by pioneering work of Filk \cite{Filk:1996dm}. Lately noncommutative coordinates appears in string theory indicating that Noncommutative Gauge Field Theory (NCGFT) could be one of its low-energy effective theories, as
proposed by Seiberg and Witten in \cite{Seiberg:1999vs}.

The noncommutativity of spacetime, for the following definition of Moyal-Weyl star($\star$)-product
\begin{equation}
(f\star g)(x)=
e^{\frac{i}{2}h\frac{\partial}{{\partial x}^\mu}
\theta^{\mu\nu}
\frac{\partial}{{\partial y}^\nu}}f(x) g(y)\bigg|_{y\to x};\:{\theta^{\mu\nu} \:\rm  constant, \:real, \:antisymmetric\: matrix,}
\label{f*g}
\end{equation}
and for local coordinates $x^\mu$ promoted to hermitian operators
$\hat x^\mu$ satisfying spacetime noncommutativity, is 
realized by  the so-called $\star$-commutator and imply uncertainty relations
\begin{equation}
[\hat x^\mu ,\hat x^\nu]=[x^\mu \stackrel{\star}{,} x^\nu]=i\theta^{\mu\nu}\Longrightarrow 
|\Delta x^\mu \Delta x^\nu| \geq \frac{1}{2}|\theta^{\mu\nu}|.
\label{x*x}
\end{equation} 

The so-called Seiberg-Witten (SW) maps~\cite{Seiberg:1999vs} and enveloping algebra based models, 
where one could deform commutative gauge theories with arbitrary gauge group and representation, having commutative instead of the noncommutative gauge symmetry preserved as the fundamental symmetry of the theory, shows a significant progress in last decade. Those are known as the Wess type of models,
see \cite{Madore:2000en,Jurco:2000ja,Jurco:2001rq,Calmet:2001na,Schupp:2001we,Jurco:2001kp, Jackiw:2001jb}.

Studies on noncommutative particle phenomenology
\cite{Trampetic:2008bk} was motivated to find possible experimental signatures and/or predict/estimate bounds on space-time noncommutativity from collider physics experimental data: for example from the Standard Model (SM) forbidden 
$Z\to\gamma\gamma$ decays, or from the SM invisible part of $Z \to \overline\nu \nu$ decays, and more important from the ultra high energy (UHE) processes occurring in the framework of the cosmogenic neutrino physics. Constraint on the scale of the NCGFT, $\Lambda_{\rm NC}$, is possible due to a direct couplings of
$Z\gamma\gamma$ and $Z \overline\nu\nu$. 

The constraints on the $\rm U(1)$ charges, known as the ``no-go theorem'' \cite{Chaichian:2009uw}, are rescinded  in our approach \cite{Horvat:2011qn}, and the noncommutative extensions of particle physics covariant SM (NCSM) and the noncommutative grand unified theories (NCGUT) models \cite{Calmet:2001na,Schupp:2001we,Jurco:2001kp,Horvat:2011qn,Behr:2002wx, Deshpande:2001mu,Duplancic:2003hg,Aschieri:2002mc,Melic:2005fm} were constructed. The method known as SW map and/or enveloping algebra (Wess approach) avoids both the gauge group and the U(1) charge issues. It was shown mathematically rigorously that any U(1) gauge theory on an arbitrary Poisson manifold can be deformation-quantized to a noncommutative gauge theory via the the enveloping algebra approach~\cite{Jurco:2001kp} and later extended to the non-Abelian gauge groups~\cite{0207arXiv0711.2965B,2009arXiv0909.4259B}. The important step that has been missed in a paper 
\cite{Chaichian:2009uw} opposing above conclusions, is the use of reducible representations \cite{Horvat:2011qn}.
All these allow a minimal deformation with no new particle content and with the sacrifice that interactions include infinitely many terms defined through recursion (due to expansions ) over the NC parameter $\theta^{\mu\nu}$.

The perturbative quantization of noncommutative field theories was first
proposed by Filk \cite{Filk:1996dm}, while
other famous examples are the running of the coupling constant of NC QED \cite{MS-R} and the UV/IR mixing \cite{Minwalla:1999px,Matusis:2000jf}. Later well behaving one-loop quantum corrections to noncommutative scalar 
$\phi^4$ theories \cite{Grosse:2004yu,arXiv:1111.5553,Magnen:2008pd} 
and the NC QED \cite{Vilar:2009er} have been found. 
Also the SW expanded NCSM \cite{Calmet:2001na,Behr:2002wx,Duplancic:2003hg,Melic:2005fm}  at first order in $\theta$, albeit breaking Lorentz symmetry is anomaly free \cite{Martin:2002nr,Brandt:2003fx}, and has well-behaved one-loop quantum corrections \cite{Minwalla:1999px,Matusis:2000jf,MS-R,Bichl:2001cq,Buric:2006wm,Latas:2007eu,Martin:2007wv,Buric:2007ix,Martin:2009vg,Martin:2009sg,Tamarit:2009iy,Buric:2010wd}.
However, despite of some significant progress in the
models \cite{Grosse:2004yu,Magnen:2008pd,arXiv:1111.5553,Vilar:2009er,Bichl:2001cq,Martin:2002nr,Brandt:2003fx,Buric:2006wm,Latas:2007eu,Martin:2007wv,Buric:2007ix,Martin:2009vg,Martin:2009sg,Tamarit:2009iy,Buric:2010wd}, a better understanding of various models quantum loop corrections still remains in general a challenging open question. This fact is particularly true for the models constructed by using SW  map expansion in the NC parameter $\theta$, \cite{Madore:2000en,Calmet:2001na,Aschieri:2002mc,Schupp:2002up,Minkowski:2003jg}. Resulting models are very useful as effective field theories including their one-loop quantum properties \cite{Bichl:2001cq,Martin:2002nr,Brandt:2003fx,Buric:2006wm,Latas:2007eu,Martin:2007wv,Buric:2007ix,Martin:2009sg,Martin:2009vg,Tamarit:2009iy,Buric:2010wd} and relevant phenomenology \cite{Melic:2005hb,Ohl:2004tn,Alboteanu:2006hh,Ohl:2010zf,Tamarit:2008vy,Horvat:2009cm,Buric:2007qx}.

Quite recently restrictions due to expansions over parameter has been overcome by constructions of the $\theta$-exact SW map and enveloping algebra based theoretical models, in the framework of covariant noncommutative quantum gauge field theory \cite{Jackiw:2001jb}, and applied in loop computation 
\cite{Schupp:2008fs,arXiv:1109.2485,arXiv:1111.4951}
and to the phenomenology, as well \cite{Horvat:2010sr,Horvat:2011iv,Horvat:2012vn}.
Namely, an expansion and cut-off in powers of 
the NC parameters $\theta^{\mu\nu}$ corresponds to an expansion in momenta  and restrict the range of validity to energies well below the NC scale $\Lambda_{\rm NC}$. Usually, this is no problem for experimental
predictions because the lower bound on the NC parameters $\theta^{\mu\nu}$
runs higher than typical momenta involved in a particular process.
However there are exotic processes, in the early universe as well as those involving ultra high energy cosmic rays
\cite{Horvat:2009cm,Horvat:2010sr,Horvat:2011iv,Horvat:2011wh},
in which the typical energy involved is higher than the current experimental
bound on the NC scale $\Lambda_{\rm NC}$.
Thus, the previous $\theta$-cut-off approximate results are inapplicable.
To cure the cut-off approximation, we are using $\theta$-exact
expressions, inspired by exact formulas for the SW
map \cite{Okawa:2001mv,Martin:2012aw}, and expand in powers of gauge fields, as we did in \cite{Horvat:2011iv}. 
In $\theta$-exact models we have studied the UV/IR mixing 
\cite{Schupp:2008fs,arXiv:1109.2485}, the neutrino propagation 
\cite{arXiv:1111.4951} and also some NC photon-neutrino phenomenology 
\cite{Horvat:2009cm,Horvat:2010sr,Horvat:2011iv,Horvat:2011wh}, respectively.  
Due to the presence of the UV/IR mixing the $\theta$-exact model 
is not perturbatively renormalizable,  thus the relations of quantum corrections to the observations \cite{Horvat:2010km} are not entirely clear. 

In this work we present NCSM extended neutrino gauge bosons actions to all orders of  $\theta$ and discuss their quantum and phenomenological properties
for light-like noncommutativity which are allowed by unitarity condition \cite{Gomis:2000zz,Aharony:2000gz}.\\

\section{Cosmogenic neutrino physics motivation}

\begin{figure}
\includegraphics[width=11cm,height=5cm]{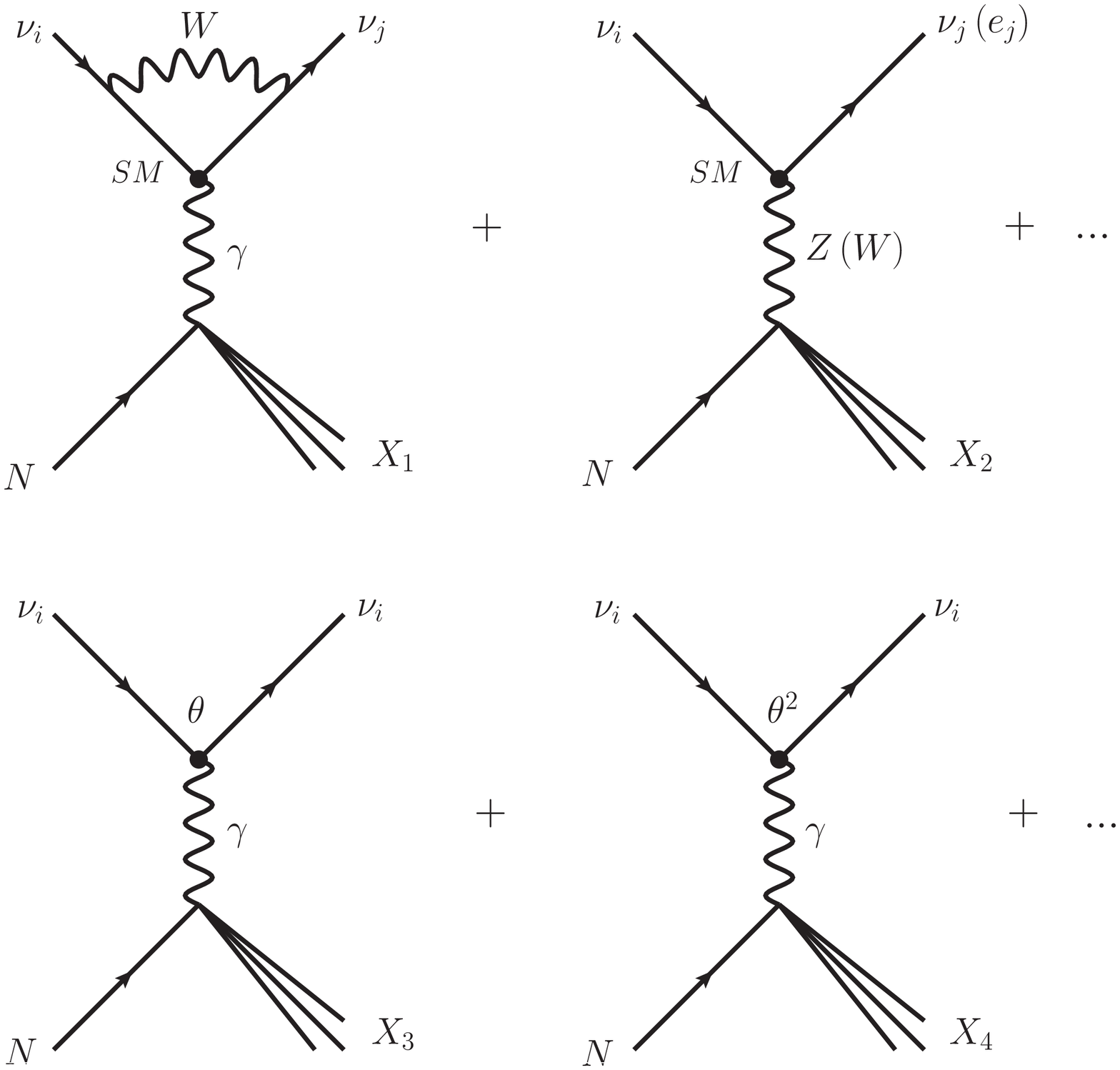}
\caption{Diagrams contributing to $\nu N\to\nu+X$ processes. }
\label{fig:trampslika}
\end{figure}

Gauge boson direct  coupling to neutral  and ``chiral'' fermion particles \cite{Schupp:2002up,Horvat:2010sr,Horvat:2011iv}, allow us via ultra-high energy cosmogenic neutrino experiments, to estimate a constraint on the scale of the NCGFT, $\Lambda_{\rm NC}$, see i.e. Fig. \ref{fig:trampslika}.
\begin{figure}
\includegraphics[width=12cm,height=5cm]{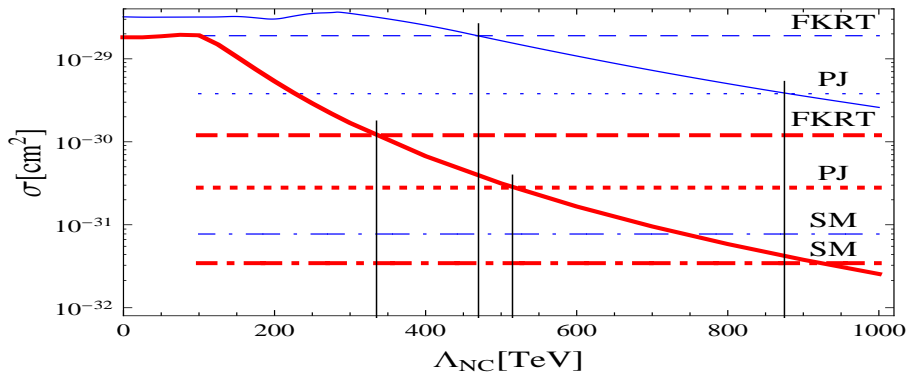}
\caption{$\nu N\to\nu\,+\, anything$ cross sections
vs. $\Lambda_{\rm NC}$ for $E_\nu=10^{10}$ GeV (thick lines) 
and $E_\nu=10^{11}$ GeV (thin lines). FKRT and PJ lines are the upper
bounds on the $\nu$-nucleon inelastic cross section, denoting different 
estimates for the cosmogenic $\nu$-flux. SM denotes the SM total
(charged current plus neutral current) $\nu$-nucleon inelastic cross
section. The vertical lines denote the intersections of our curves with the
RICE results.  }
\label{fig:ncSM-CrossSections}
\end{figure}
\begin{figure}
\includegraphics[width=12cm,height=5cm]{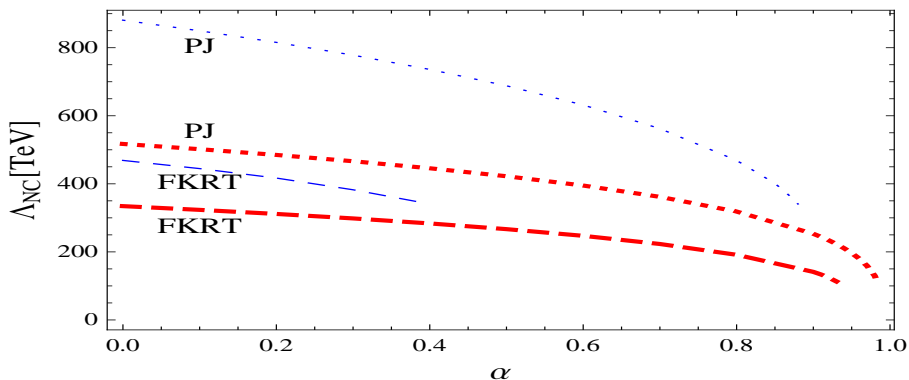}
\caption{The intersections of our curves with the RICE results (cf. Fig.2)
as a function of the fraction of Fe nuclei in the UHE cosmic rays. The
terminal point on each curve represents the highest fraction of Fe nuclei
above which no useful information on $\Lambda_{\rm NC}$ can be inferred with
our method.}
\label{fig:ncSM-LambdaVsAlpha}
\end{figure}

The observation of ultra-high energy (UHE) $\nu$'s from extraterrestrial
sources would open a new window to look to the cosmos, as such $\nu$'s may
easily escape very dense material backgrounds around local astrophysical
objects, giving thereby information on regions that are otherwise hidden to
any other means of exploration. In addition, $\nu$'s are not deflected on
their way to the earth by various magnetic fields, pointing thus back to
the direction of distant UHE cosmic-ray source candidates. This could also
help resolving the underlying acceleration in astrophysical sources.
     
In the energy spectrum of UHE cosmic rays at $\sim 4 \times 10^{19}$ eV the GZK-structure has been observed recently with high statistical accuracy \cite{Aglietta:2007yx}. Thus the flux of the so-called 
cosmogenic $\nu$'s, arising from photo-pion production on the
cosmic microwave background $p \gamma_{CMB} \rightarrow \Delta^{*}
\rightarrow N\pi$ and subsequent pion decay, is now guaranteed to exist. 
Possible ranges for the size of the flux of cosmogenic $\nu$'s can be obtained from separate analysis of the data  \cite{Fodor:2003ph,Protheroe:1995ft}. 

Using the upper bound on the $\nu N$ cross section derived from the RICE
Collaboration search results \cite{Kravchenko:2002mm} at $E_{\nu} =
10^{11}$ GeV ($4\times 10^{-3}$ mb for the FKRT $\nu$-flux \cite{Fodor:2003ph})), 
one can infer from $\theta$-truncated model    
on the NC scale $\Lambda_{\rm NC}$ to be greater than 455 TeV, 
a really strong bound. Here we use $\theta^{\mu\nu} \equiv c^{\mu\nu}/\Lambda_{\rm NC}^2 $ such that 
the matrix elements of $c^{\mu\nu}$ are of order 1.

One should however be
careful and suspect this result as it has been obtained from the conjecture
that the $\theta$-expansion stays well-defined in the kinematical region of interest. 
Although a heuristic criterion for the validity of the perturbative $\theta$-expansion,
$\sqrt{s}/\Lambda_{\rm NC} \lesssim\;1$, with $s = 2 E_{\nu}M_N$, would
underpin our result on $\Lambda_{\rm NC}$, a more thorough inspection on
the kinematics of the process does reveal a  more stronger energy
dependence  $E_{\nu}^{1/2} s^{1/4}/ \Lambda_{\rm NC} \lesssim 1$.
In spite of an additional phase-space suppression for small $x$'s in  
the $\theta^2$-contribution \cite{Ohl:2004tn} of
the cross section relative to the $\theta$-contribution, we find an unacceptably 
large ratio $\sigma({\theta^2})/\sigma({\theta}) \simeq 10^4$,
at $\Lambda_{\rm NC}=455$ TeV.
Hence, the bound on $\Lambda_{\rm NC}$ obtained this way is incorrect, 
and our last resort 
is to modify the model adequately to include the full-$\theta$
resummation, thereby allowing us to compute nonperturbatively in $\theta$.
Total cross section, as a function of the NC scale at fixed $E_{\nu} = 10^{10}$ GeV 
and $E_{\nu} = 10^{11}$ GeV, together with the upper bounds depending on the actual size of the cosmogenic $\nu$-flux (FKRT \cite{Fodor:2003ph} and PJ \cite{Protheroe:1995ft}) as well as the total SM cross sections at these energies, are depicted in our Figure \ref{fig:ncSM-CrossSections}.

Even if the future data confirm that UHE cosmic rays are composed mainly of Fe
nuclei, as indicated by the PAO data, then still
valuable information on $\Lambda_{\rm NC}$ can be obtained with our method, as
seen in Fig.\ref{fig:ncSM-LambdaVsAlpha}. 
Here we see the intersections of our curves with the
RICE results (cf. Fig.\ref{fig:ncSM-CrossSections})
as a function of the fraction $\alpha$ of Fe nuclei in the UHE cosmic rays.
Results depicted in Figs.\ref{fig:ncSM-CrossSections}-\ref{fig:ncSM-LambdaVsAlpha},
shows convergent behavior.  In our opinion those were the strong signs   
to continue research towards  quantum properties and phenomenology of 
such $\theta$-exact NCGFT model. 

\section{Covariant $\theta$-exact ${\rm U_{\star}(1)}$ model}

We start with the following SW type of NC $\rm U_{\star}(1)$ gauge model:
\begin{equation}
S=\int-\frac{1}{4}F^{\mu\nu}\star F_{\mu\nu}+i\overline\Psi\star\fmslash{D}\Psi\,,
\label{S}
\end{equation}
with the NC definitions of the nonabelian field strength and the covariant derivative, respectively:
\begin{eqnarray}
F_{\mu\nu}&=&\partial_\mu A_\nu-\partial_\nu
A_\mu-i[A_\mu\stackrel{\star}{,}A_\nu],\;\;
D_\mu\Psi=\partial_\mu\Psi-i[A_\mu\stackrel{\star}{,}\Psi].
\label{DF}
\end{eqnarray}
All noncommutative fields in this action $(A_\mu,\Psi)$ are images under (hybrid) Seiberg-Witten maps
of the corresponding commutative fields $(a_\mu,\psi)$.
Here we shall interpret the NC fields as valued in
the enveloping algebra of the underlying gauge group.
This naturally corresponds to an expansion in powers of
the gauge field $a_\mu$ and hence in powers of
the coupling constant $e$. At each order in $a_\mu$ we shall
determine $\theta$-exact expressions.

In the next step we expand the action in terms of
the commutative gauge fields $a_\mu$ (and/or coupling constant)
 and using the SW map solution \cite{Schupp:2008fs}
up to the $\mathcal O(a^3)$ order:
\begin{eqnarray}
\Lambda&=&\lambda-\frac{1}{2}\theta^{ij}a_i\star_2\partial_j\lambda, \;\;
A_\mu=\,a_\mu-\frac{1}{2}\theta^{\nu\rho}{a_\nu}\star_2(\partial_\rho
a_\mu+f_{\rho\mu}),
\nonumber\\
\Psi&=&\psi-\theta^{\mu\nu}
{a_\mu}\star_2{\partial_\nu}\psi+\frac{1}{2}\theta^{\mu\nu}\theta^{\rho\sigma}
\bigg[(a_\rho\star_2(\partial_\sigma
a_\mu+f_{\sigma\mu}))\star_2{\partial_\nu}\psi
\nonumber\\
&+&2a_\mu{\star_2}
(\partial_\nu(a_\rho{\star_2}\partial_\sigma\psi))
-a_\mu{\star_2}(\partial_\rho
a_\nu{\star_2}\partial_\sigma\psi)-\big(a_\rho\partial_\mu\psi(\partial_\nu
a_\sigma+f_{\nu\sigma})
-\partial_\rho\partial_\mu\psi a_\nu
a_\sigma\big)_{\star_3}\bigg],
\label{SWmap}
\end{eqnarray}
with $\Lambda$ being the NC gauge parameter and
$f_{\mu\nu}$ is the abelian commutative field strength
$f_{\mu\nu}=\partial_\mu a_\nu-\partial_\nu a_\mu$.

The generalized Mojal-Weyl star products
$\star_2$ and $\star_3$, appearing in (\ref{SWmap}), are defined, respectively, as
\begin{eqnarray}
f(x)\star_2 g(x)&=&[f(x) \stackrel{\star}{,}g(x)]
=\frac{\sin\frac{\partial_1\theta
\partial_2}{2}}{\frac{\partial_1\theta
\partial_2}{2}}f(x_1)g(x_2)\bigg|_{x_1=x_2=x}\,,
\label{f*2g}
\end{eqnarray}
\begin{eqnarray}
(f(x)g(x)h(x))_{\star_3}&=&\Bigg(\frac{\sin(\frac{\partial_2\theta
\partial_3}{2})\sin(\frac{\partial_1\theta(\partial_2+\partial_3)}{2})}
{\frac{(\partial_1+\partial_2)\theta \partial_3}{2}
\frac{\partial_1\theta(\partial_2+\partial_3)}{2}}
+\{1\leftrightarrow 2\}\,\Bigg)f(x_1)g(x_2)h(x_3)\bigg|_{x_i=x},
\nonumber\\
\label{fgh*3}
\end{eqnarray}
where $\star$ is associative but noncommutative, while $\star_2$ and $\star_3$
are both commutative but nonassociative. 

\begin{figure}
\includegraphics[width=12cm,height=4cm]{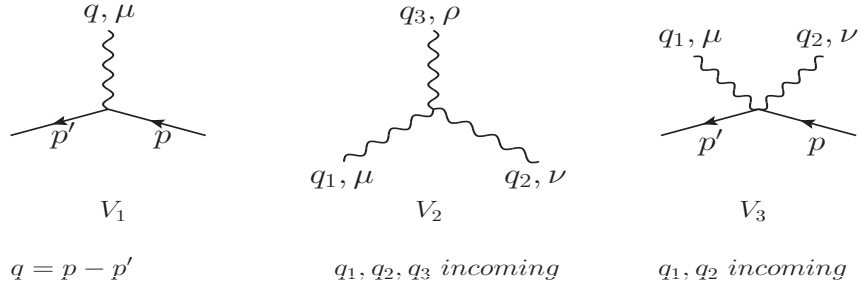}
\caption{Three- and your-field vertices}
\label{fig:Bayzell11Vert}
\end{figure}

The resulting expansion defines $\theta$-exact neutrino-photon ${\rm U_{\star}(1)}$ 
actions, for a gauge and a matter sectors respectively. Pure gauge field (3-photon) action reads:
\begin{eqnarray}
S_g&=&\int \;i\partial_\mu a_\nu\star[a^\mu\stackrel{\star}{,}a^\nu]
+\frac{1}{2}\partial_\mu
\bigg(\theta^{\rho\sigma}a_\rho\star_2(\partial_\sigma a_{\nu}+f_{\sigma\nu})\bigg)\star
f^{\mu\nu}.
\label{Sgauge}
\end{eqnarray}
The photon-fermion action up to 2-photon 2-neutrino fields can be derived
by using the first order gauge field and the second order neutrino field expansions,
\begin{eqnarray}
S_f&=&\int \;\bigg(\overline\psi
+(\theta^{ij}\partial_i\overline\psi \star_2
a_j)\bigg)\gamma^\mu[a_\mu\stackrel{\star}{,}\psi]
+ i(\theta^{ij}\partial_i\overline\psi
\star_2 a_j)\fmslash\partial\psi-i\overline\psi\star
\fmslash\partial(\theta^{ij}
a_i\star_2\partial_j\psi)
\nonumber\\
&-&
\!\overline\psi\gamma^\mu[a_\mu\!\stackrel{\star}{,}\!\theta^{ij}
a_i\!\star_2\!\partial_j\psi]\!-
\!\overline\psi\gamma^\mu
\bigg[\frac{1}{2}\theta^{ij}a_i\!\star_2\!(\partial_j
a_\mu\!+\!f_{j\mu})\!\stackrel{\star}{,}\!\psi\bigg]\!
-
\!i(\theta^{ij}\partial_i\overline\psi
\!\star_2\!a_j)\fmslash\partial(\theta^{kl}
a_k\!\star_2\!\partial_l\psi)
\nonumber\\
&+&
\frac{i}{2}\theta^{ij}\theta^{kl}
\bigg[(a_k\star_2(\partial_l
a_i+f_{li}))\star_2\partial_j\overline\psi
+
2a_i\star_2(\partial_j(a_k\star_2\partial_l\overline\psi))-
a_i\star_2(\partial_k
a_j\star_2\partial_l\overline\psi)
\nonumber\\
&+&
\big(a_i\partial_k\overline\psi(\partial_j
a_l+f_{jl})-\partial_k\partial_i\overline\psi a_j a_l\big)_{\star_3}\bigg]
\fmslash\partial\psi
+
\frac{i}{2}\theta^{ij}\theta^{kl}\overline\psi\fmslash\partial
\bigg[(a_k\star_2(\partial_l
a_i+f_{li}))\star_2\partial_j\psi
\nonumber\\
&+&
\!2a_i\!\star_2\!
(\partial_j(a_k\!\star_2\!\partial_l\psi))\!-\!a_i\!\star_2\!(\partial_k
a_j\!\star_2\!\partial_l\psi)\!
+
\big(a_i\partial_k\psi(\partial_j
a_l\!+\!f_{jl})\!-\!\partial_k\partial_i\psi a_j
a_l\big)_{\star_3}\bigg]\,.
\label{Sfermion}
\end{eqnarray}
Note that actions for gauge and matter fields obtained above,
(\ref{Sgauge}) and (\ref{Sfermion}) respectively, are nonlocal
objects due to the presence of the star products: $\star$, $\star_2$ and $\star_3$.
Feynman rules from above actions (Fig.\ref{fig:Bayzell11Vert}), are given explicitly in \cite{arXiv:1111.4951}.

\section{Quantum properties: neutrino self energy}

\begin{figure}
\includegraphics[width=12cm,height=6cm]{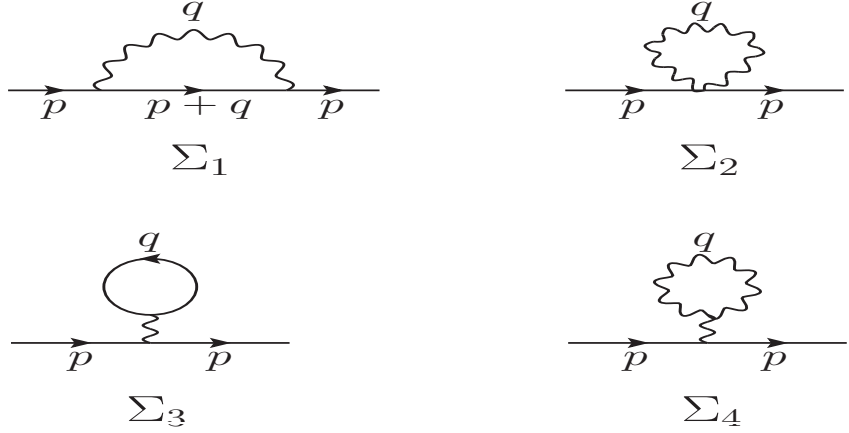}
\caption{One-loop self-energy of a massless neutrino}
\label{Sigma1-loop}
\end{figure}

As depicted in Fig. \ref{Sigma1-loop}, there are four Feynman diagrams
contributing  to the  $\nu$-self-energy at one-loop.
With the aid of (\ref{Sfermion}), 
we have verified by explicit calculation that the 4-field
tadpole ($\Sigma_2$) does vanish. The 3-fields tadpoles ($\Sigma_3$ 
and $\Sigma_4$) can be ruled out by invoking
the NC charge conjugation symmetry \cite{Aschieri:2002mc}.
Thus only the $\Sigma_1$ diagram needs to be evaluated.   
In spacetime of the dimensionality $D$ we obtain
\begin{eqnarray}
\Sigma_1
&=&\mu^{4-D}\int \frac{d^D q}{(2\pi)^D}
\bigg(\frac{\sin\frac{q\theta p}{2}}{\frac{q\theta
p}{2}}\bigg)^2\frac{1}{q^2}\frac{1}{(p+q)^2}
\label{Sigma1}\\
&\cdot & \Bigg[(q\theta p)^2(4-D)(\fmslash p+\fmslash q)
+(q\theta p)\bigg(\fmslash{\tilde q}(2p^2+2p\cdot q)-\fmslash{\tilde p}(2q^2+2p\cdot q)\bigg)
\nonumber\\
&+&\fmslash p\bigg(\tilde q^2(p^2+2p\cdot q)-q^2(\tilde p^2+2\tilde
p\cdot\tilde q) \bigg)
+
\fmslash q\bigg(\tilde p^2(q^2+2p\cdot q)-p^2(\tilde
q^2+2\tilde p\cdot\tilde q )\bigg)\Bigg]\,,
\nonumber
\end{eqnarray}
where ${\tilde p}^\mu=(\theta p)^\mu=\theta^{\mu\nu} p_\nu $,
and in addition ${\tilde{\tilde p}}^\mu=(\theta\theta p)^\mu
=\theta^{\mu\nu}\theta_{\nu\rho}p^\rho$.
To perform computations of those integrals using the dimensional
regularization method, we first use
the Feynman parametrization on the quadratic denominators,
then the Heavy Quark Effective theory (HQET)
parametrization \cite{Grozin:2000cm} is used to combine
the quadratic and linear denominators.
In the next stage we use the Schwinger
parametrization to turn the denominators
into Gaussian integrals. Evaluating the relevant integrals for
$D=4-\epsilon$ in the limit $\epsilon\to 0$, we obtain the 
closed form self-energy
\begin{eqnarray}
\Sigma_1&=&\gamma_{\mu}
\bigg[p^{\mu}\: A+(\theta{\theta p})^{\mu}\;\frac{p^2}{(\theta p)^2}\;B\bigg]\,,
\label{sigma1AB}\\
A&=& \frac{-1}{(4\pi)^2}
\bigg[p^2\;\bigg(\frac{\tr\theta\theta}{(\theta p)^2}
+2\frac{(\theta\theta p)^2}{(\theta p)^4}\bigg) A_1 
+
 \bigg(1+p^2\;\bigg(\frac{\tr\theta\theta}{(\theta p)^2}
+\frac{(\theta\theta p)^2}{(\theta p)^4}\bigg)\bigg)A_2 \bigg]\,,
\label{A}\\
A_1&=&\frac{2}{\epsilon}+\ln(\mu^2(\theta p)^2)
+ \ln({\pi e^{\gamma_{\rm E}}})
+
\sum\limits_{k=1}^\infty \frac{\left(p^2(\theta p)^2/4\right)^k}{\Gamma(2k+2)}
\left(\ln\frac{p^2(\theta p)^2}{4} + 2\psi_0(2k+2)\right) \,,
\label{A1}\\
A_2&=&-\frac{(4\pi)^2}{2} B = -2
\nonumber\\
&+& \sum\limits_{k=0}^\infty
\frac{\left(p^2(\theta p)^2/4\right)^{k+1}}{(2k+1)(2k+3)\Gamma(2k+2)}
\Bigg(\ln\frac{p^2(\theta p)^2}{4} 
-
2\psi_0(2k+2)
- \frac{8(k+1)}{(2k+1)(2k+3)} \Bigg),
\label{A2}
\end{eqnarray}
with  $\gamma_{\rm E}\simeq0.577216$ being Euler's constant.

The $1/\epsilon$ UV divergence
could in principle be removed by a properly chosen counterterm.
However due to the specific momentum-dependent 
coefficient in front of it, a nonlocal form for it is required. 

Turning to the UV/IR mixing problem, we recognize
a soft UV/IR mixing term represented by a logarithm,
\begin{equation}
\Sigma_{\rm UV/IR}={\fmslash p} \:\frac{p^2}{(4\pi)^2}
\left(\ln\frac{1}{|\mu(\theta p)|^2}\right)
\bigg(\frac{\tr\theta\theta}{(\theta p)^2}
+2\frac{(\theta\theta p)^2}{(\theta p)^4}\bigg).
\label{lnUV/IR}
\end{equation}
Instead of dealing with nonlocal counterterms, we take a different route
here to  cope with various divergences besetting (\ref{sigma1AB}). Since $\theta^{0i}
\neq 0$ makes a NC theory nonunitary,
we can without loss of generality
chose $\theta$ to lie in the (1, 2) plane
\begin{equation}
\theta^{\mu\nu}=\frac{1}{\Lambda_{\rm NC}^2}
\begin{pmatrix}
0&0&0&0\\
0&0&1&0\\
0&-1&0&0\\
0&0&0&0
\end{pmatrix}
\:\:\longrightarrow \:\:
\frac{\tr\theta\theta}{(\theta p)^2}
+2\frac{(\theta\theta p)^2}{(\theta p)^4}=0,\:\forall p.
\label{degen}
\end{equation}
With (\ref{degen}), $\Sigma_1$,
in terms of Euclidean momenta, receives the following form:
\begin{equation}
\Sigma_1=\frac{-1}{(4\pi)^2}\gamma_\mu\left[p^\mu
\bigg(1 + \frac{\tr\theta\theta}{2}\frac{p^2}{(\theta p)^2}\bigg)
-2(\theta\theta p)^\mu\frac{p^2}{(\theta p)^2} \right] A_2.
\label{Sigmabuble}
\end{equation}
By inspecting (\ref{A2}) one can be easily convinced that $A_2$ is 
free from the $1/\epsilon$ divergence and the UV/IR mixing term, being also 
well-behaved in the infrared, in the $\theta \rightarrow 0$ 
as well as $\theta p \rightarrow 0$
limit. We see, however, that the two terms in (\ref{Sigmabuble}), 
one being proportional to
$\fmslash{p}$ and the other proportional to $\fmslash{\tilde{\tilde p}}$,
are still ill-behaved in the $\theta p \rightarrow 0$
limit. If, for the choice (\ref{degen}), $P$ denotes the momentum in the (1, 2)
plane, then $\theta p = \theta P$. For instance, a particle moving
inside the NC plane with
momentum $P$ along the one axis, has a spatial extension of size $|\theta P|$
along the other. For the choice (\ref{degen}), $\theta p \rightarrow 0$ corresponds to a zero momentum projection onto the (1, 2) plane. Thus, albeit in our approach the commutative limit ($\theta \rightarrow 0$) is smooth at the quantum level,
the limit when an extended object (arising due to the fuzziness of space)
shrinks to zero, is not. We could surely claim that in our approach the
UV/IR mixing problem is considerably softened; on the other hand, we have
witnessed how the problem strikes back in an unexpected way. This is, at the
same time, the first example where this two limits are not degenerate.

Computing dispersion relations we probe physical consequence of
the 1-loop quantum correction, with $\Sigma_{1-loop_{M}}$ 
from Eq. (3.25) in \cite{arXiv:1111.4951}. We have to modify the propagator
\begin{equation}
\frac{1}{\fmslash\Sigma}=\frac{1}{\fmslash p-\Sigma_{1-loop_{M}}}=
\frac{\fmslash\Sigma}{\Sigma^2}\,,
\label{disp}
\end{equation}
and further we choose the NC parameter to be (\ref{degen})
so that the denominator is finite and can be expressed explicitly:
\begin{equation}
\hspace{-.05cm}\Sigma^2=p^2\left[\hat A_2^2\left(\frac{p^4}{p^4_r}
+2\frac{p^2}{p_r^2}+5\right)-\hat A_2\left(6+2\frac{p^2}{p_r^2}\right)+1\right],\,
\label{23}
\end{equation}
where $p_r$ represents $r$-component of the momentum $p$
in a cylindrical spatial coordinate system and $\hat A_2= e^2A_2/(4\pi)^2=-B/2$.

From above one see that $p^2=0$ defines one set of the dispersion relation,
corresponding to the dispersion for the massless neutrino mode,
however the denominator $\Sigma^2$ has one more coefficient $\Sigma'$
which could also induce certain zero-points. Since the $\hat A_2$ is a
function of a single variable $p^2p_r^2$, with $
p^2=p^2_0-p^2_1-p^2_2-p^2_3\,\,\rm and\,\,p^2_r=p^2_1+p^2_2$,
the condition $\Sigma'=0$  can be expressed as a simple algebraic equation
\begin{equation}
\hat A_2^2z^2-2\left(A_2-\hat A_2^2\right)z
+\left(1-6\hat A_2+5\hat A_2^2\right)=0\,,
\label{A2z2}
\end{equation}
of new variables $z:=p^2/p_r^2$, in which the coefficients
are all functions of $y:=p^2p^2_r/\Lambda^4_{\rm NC}$.
Formal solutions of (\ref{A2z2})
\begin{equation}
z=\frac{1}{\hat A_2}\left[\left(1-\hat A_2\right)\pm\,2\left(\hat A_2
-\hat A_2^2\right)^{\frac{1}{2}}\right]\,,
\label{zsolutions}
\end{equation}
are birefringent.
The behavior of solutions (\ref{zsolutions}), is next analyzed at
two limits $y\to \,0$, and $y\to\,\infty$.

\subsubsection{The low-energy regime: $p^2p^2_r \ll\Lambda^4_{\rm NC}$}

For $y \ll 1$ we set $\hat A_2$ to its zeroth order value $ e^2/8\pi^2$, 
\begin{eqnarray}
p^2&\sim& \left(\left(\frac{8\pi^2}{e^2}-1\right)\pm 2
\left(\frac{8\pi^2}{e^2}-1\right)^{\frac{1}{2}}\right)\cdot\,p_r^2
\simeq\left(859\pm 59\right)\cdot\,p_r^2\,,
\label{859pm59}
\end{eqnarray}
obtaining two (approximate) zero points. From the definition of $p^2$
and $p_r^2$ we see that both solutions are real and positive.
Taking into account the higher order (in y) correction
these poles will locate nearby the real axis of
the complex $p_0$ plane thus correspond to some metastable modes with
the above defined dispersion relations. As we can see, 
the modified dispersion relation \eqref{859pm59} does not depend on 
the noncommutative scale, therefore it introduces a discontinuity in 
the $\Lambda_{\rm NC}\to\infty$ limit, 
which is not unfamiliar in noncommutative theories.

\subsubsection{The high-energy regime: $p^2p^2_r \gg \Lambda^4_{\rm NC}$}
At $y\gg 1$ we analyze the asymptotic behavior of $A_2$, therefore \eqref{zsolutions} can be reduced
\begin{equation}
A_2\sim \frac{i\pi^2}{8} {\sqrt y}\left(1-\frac{16i}{\pi y}
e^{-\frac{i}{2}{\sqrt y}}\right)+\mathcal O\left(y^{-1}\right),
\;\; \longrightarrow \;\;
z\sim -1\pm 2i \;\; \rightarrow \;\;p^2_0\sim p^2_3\pm 2i p^2_r.
\end{equation}
We thus reach two unstable deformed modes besides the usual mode $p^2=0$ in the high energy regime. Here again the leading order deformed dispersion relation does not depend on the noncommutative scale $\Lambda_{\rm NC}$.

\section{Phenomenology: Rate of $Z\to \overline\nu\nu$ decays}

To illustrate phenomenologicall effects of our $\theta$-exact construction, 
we present a computation the $Z\to\overline\nu\nu$ decay rate in 
the Z--boson rest frame, which is then readily to be compared with 
the precision Z resonance measurements, where Z is almost at rest.
Since the complete $Z\overline\nu\nu$ interaction on noncommutative spaces was discussed in details in \cite{Horvat:2011qn,Horvat:2011iv,arXiv:1109.2485,arXiv:1111.4951},  we shall not repeat it here. Using the {\it almost complete} 
$Z\overline\nu\nu$ vertex from \cite{Horvat:2011qn} we have found the following $Z\to \overline\nu\nu$ partial width \cite{Horvat:2012vn}
\begin{eqnarray}
\Gamma(Z\to\overline\nu\nu)&=&
\Gamma_{\rm SM}(Z\to\overline\nu\nu)
\nonumber\\
&+&\frac{\alpha}{3 M_Z |\vec{E_\theta}|}
   \Bigg[\kappa  \left(1 -\kappa +\kappa\cos 2 \theta _W \right) 
   \sec ^2\theta_W\cos\left(\frac{M_Z^2 |\vec{E_\theta}|}{4}\right)-8 \csc ^2 2 \theta _W \Bigg] 
   \sin\left(\frac{M_Z^2 |\vec{E_\theta}|}{4}\right)
\nonumber\\
&+&\frac{\alpha M_Z}{12}   \bigg[-2 \kappa ^2
+(\kappa  (2 \kappa -1)+2) \sec^2\theta _W+2 \csc^2\theta _W\bigg],
\label{rateZnunu}
\end{eqnarray}
where $\kappa$ is an arbitrary constant\footnote{The constant $\kappa$ measures a correction from the  $\star$-commutator coupling of the right handed neutrino $\nu_R$ to the noncommutative hypercharge $\rm U_{\star}(1)_Y$ gauge field $B^0_\mu[\kappa]$. Coupling is chiral blind and it vanishes in the commutative limit. The non-$\kappa$-proportional term, on the other hand, is the noncommutative deformation of standard model Z-neutrino coupling, which involves the left handed neutrinos only. Details can be found in section four of \cite{Horvat:2011qn}. }.
The NC part vanishes when $\vec{E}_\theta\to 0$, i.e. for vanishing $\theta$ or space-like noncommutativity, but not light-like \cite{Gomis:2000zz,Aharony:2000gz}.

\begin{figure}[top]
\includegraphics[width=12cm,height=5cm]{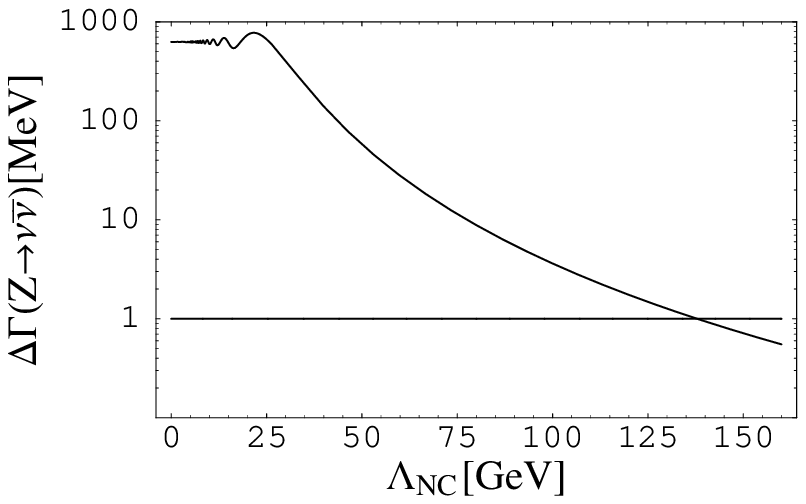}
\caption{$\Delta\Gamma$ decay width vs. $\Lambda_{\rm{NC}}$.}
\label{fig:Z2nubarnuNC}
\end{figure}
A comparison of the experimental Z decay width $\Gamma_{\rm invisible}=(499.0\pm 1.5)$ MeV \cite{PDG2011} with its SM theoretical counterpart, allows us to set a constraint $\Gamma(Z\to\overline\nu\nu) - \Gamma_{\rm SM}(Z\to\overline\nu\nu) \lesssim 1$ MeV, from where a bound on the scale of noncommutativity
$\Lambda_{\rm{NC}} = {|\vec{E_\theta}|^{-1/2}}\stackrel{>}\sim 140$ GeV is obtained  (see Fig.~\ref{fig:Z2nubarnuNC}), for the choice $\kappa =1$.

\section{Conclusions}

In the energy range of interest, $10^{10}$ to ${10^{11}}$ GeV,  where there is always energy of the system $(E)$ larger than the NC scale 
$(E/\Lambda_{\rm NC}>1)$, the perturbative expansion in terms of $\Lambda_{\rm NC}$ retains no longer its meaningful character, thus it is
forcing us to resort to those NC field-theoretical frameworks involving the 
full $\theta$-resummation. Our numerical estimates of the contribution to the processes coming from the photon exchange, pins impeccably down a lower bound on $\Lambda_{\rm NC}$ to be as high as around up to ${\cal O}(10^6)$ GeV,
depending on the cosmogenic $\nu$-flux. 

We first discuss $\theta$-exact computation of the
one-loop quantum correction to the $\nu$-propagator. 
General expression for the neutrino self-energy
(\ref{sigma1AB}) contains in (\ref{A1}) both a hard $1/\epsilon$
UV term and  the UV/IR mixing term
with a logarithmic infrared singularity $\ln |\theta p|$.
Results shows complete decoupling of the UV divergent term 
from softened UV/IR mixing term and from the finite terms as well. 
Our deformed dispersion relations at both the low and high energies 
and at the leading order
do not depend on the noncommutative scale $\Lambda_{\rm NC}$.
The low energy dispersion \eqref{859pm59} is capable of generating a direction dependent superluminal velocity. This is clear from the maximal attainable velocity of the neutrinos
\begin{equation}
\frac{{\rm v}_{max}}{c}=\frac{dE}{d|\vec p|}\sim \sqrt{
1+\left(859\pm 59\right)
\sin^2\vartheta}\,,
\label{varepsilon}
\end{equation}
where $\vartheta$ is the angle with respect to the direction perpendicular 
to the NC plane. This gives one more example how such spontaneous $\theta$-background breaking of Lorentz symmetry could affect  the particle kinematics through quantum corrections, even without divergent behavior like UV/IR mixing. 
On the other hand one can also see that the magnitude of superluminosity 
is in general very large in our model as a quantum effect, thus seems contradicting various observations which suggests much smaller values 
\cite{Hirata:1987hu,Bionta:1987qt,Longo:1987ub}.   
On the other hand, note that the large superluminal velocity issue may also be reduced/removed by taking into account several considerations and/or properties:\\
- Selection of a constant nonzero $\theta$ 
background in this paper is due to the computational simplicity. 
The results will, however, still hold for a NC background 
that is varying sufficiently slowly with respect to the scale of
noncommutativity. There is no physics reason to expect 
$\theta$ to be a globally constant background {\it ether}.  
In fact, if the $\theta$ background is only nonzero in 
tiny regions (NC bubbles) the effects of the modified dispersion
relation will be suppressed macroscopically.
Certainly a better understanding of possible sources of NC is needed.\\
- We have considered only the purely noncommutative 
neutrino-photon coupling. However, it has been pointed out that modified 
neutrino dispersion relation could open decay channels within 
the commutative standard model framework \cite{Cohen:2011hx}. 
In our case this would further provide cascade decay channel(s) which can, 
via bremsstrahlung, bring superluminal neutrinos to normal ones.\\
- Our results differs with respect to non SW map models
since in our case both terms are proportional to the spacetime
noncommutativity dependent $\theta$-ratio (the scale-independent structure!) 
factor in (\ref{degen}), which arise from the natural non-locality of our actions.
Besides the divergent terms, a new spinor structure $(\theta\theta p)$
with finite coefficients emerges in our computation, see (\ref{sigma1AB})-(\ref{A2}).
All these structures are proportional to $p^2$, therefore if appropriate
renormalization conditions are imposed, the commutative dispersion relation
$p^2=0$ can still hold.
Also it is reasonable to conjecture that SW map freedom may also serve as one possible 
remedy to (23) issue.\\
- Finally, we mention that our approach to UV/IR mixing should not be confused 
with the one  based on a theory with UV completion ($\Lambda_{\rm UV} < \infty$), 
where a theory becomes an effective QFT, and the UV/IR mixing manifests itself via 
a specific relationship between the UV and the IR cutoffs
\cite{AlvarezGaume:2003mb}.
This is fortunate with regard to the use of  low-energy NCQFT
as an important window  to holography \cite{Horvat:2010km} and quantum gravity \cite{Szabo:2009tn}.



\section*{Acknowledgment}
Work supported by the Croatian Ministry of Science, Education and
Sport project 098-0982930-2872. I would like to thank R. Horvat, D. Kekez, A. Ilakovac and J. You for many
valuable comments/remarks. 



\end{document}